\documentstyle[preprint,aps,epsfig]{revtex}  
\tightenlines 
\begin{document}    
\draft     
\title{Realistic extremely flat scalar potential in 3-3-1 models} 
\author{Alex G. Dias} 
\address{Instituto de F\'\i sica, Universidade de S\~ao Paulo, \\ 
C. P. 66.318, 05315-970\\ 
S\~ao Paulo, SP\\ Brazil}  
\author{V. Pleitez} 
\address{Instituto de F\'\i sica Te\'orica, Universidade Estadual Paulista,\\ 
Rua Pamplona 145, \\ 
01405-900 S\~ao Paulo, SP \\
Brazil}  
\author{M. D. Tonasse}   
\address{Instituto Tecnol\'ogico de Aeron\'autica, Centro T\'ecnico  
Aeroespacial\\  
Pra\c ca Marechal do Ar Eduardo Gomes 50, 12228-901 \\ 
S\~ao Jos\'e  dos Campos, SP\\ 
Brazil}   
\date{\today}   
\maketitle     
\begin{abstract}   
We show that in 3-3-1 models it is possible to implement an extremely flat
scalar potential, i.e., a zero contribution to the
cosmological constant, and still having realistic values for the masses of the
scalar Higgs fields. Besides, when loop corrections are considered they impose
constraints over heavy particles masses (exotic quarks and extra
vector bosons) which are present in the model. 
A crucial ingredient in 3-3-1 models is the existence of trilinear terms in the
scalar potential. We also consider the two-Higgs doublets
extension of the standard model with and without supersymmetry.  
\end{abstract} 

\section{Introduction} 
\label{sec:intro}  

The cosmological constant $\Lambda$ was initially postulated by Einstein who
included by hand the term $\Lambda g_{\mu\nu}$ in his gravitational theory in
order to obtain a static Universe solution. In subsequent works the meaning of
this term was clarified and it was shown that it is linked with the vacuum
energy density. The Einstein's field equations with the $\Lambda$-term are  
\begin{equation} 
R_{\mu\nu} - \frac{R}{2}g_{\mu\nu} - \Lambda g_{\mu\nu} = 8\pi GT_{\mu\nu}, 
\label{eq}
\end{equation}  
where $R_{\mu\nu}$ is the Ricci tensor, $g_{\mu\nu}$ is the metric tensor and
$R$ represents the curvature scalar of the space-time and $T_{\mu\nu}$ is the
energy momentum tensor; $G$ is the constant of gravitation.
The so-called ``cosmological constant
problem''~\cite{WE89} consists in the fact that the values of the vacuum
energy densities related to quantum field energy scales, for instance, quark
condensates $\langle q\bar{q}\rangle$, $\Lambda_{\rm QCD}$ and to the Planck
scale $\Lambda_{\rm Planck}$, are several orders of magnitude larger than the
values suggested by the astronomical observations (see below).    

In the context of classical general relativity we can always omit the
$\Lambda$-term in the Lagrangian. On the other hand, in quantum field theory
only energy differences are measured, so we can always redefine the vacuum
energy in such a way that it always corresponds to zero energy. 
However, since gravitation is sensible to the absolute value of the
energy, through distortions in the space-time, all matter and energy forms
couple to gravitation. It means that the metric $g_{\mu\nu}$ can be coupled as
an external field to the bare action of a quantum field theory playing the role
of a source for the energy-momentum operator~\cite{ue}. Therefore we cannot
avoid the problem when both gravitational phenomena and quantum field theory are
taken into account, because in this situation it is not allowed to redefine the
vacuum energy~\cite{ZE68,CS01,indio}.   

Until the end of the nineties, astronomical data were able to give for the value
of $\vert\Lambda_{\rm obs}\vert$ only an upper bound. Because of the smallness
of the bound, compared with the theoretically expected values, many  attempts
were made in order to find models in which the total cosmological constant was
exactly zero~\cite{DO97}. None of these attempts, however, is based on some
fundamental theory and at present only considerations based on the
anthropic principle suggest a route towards the response to the cosmological
constant question. However, although these considerations receive support from 
the inflationary cosmological models, they do not have yet experimental
basis~\cite{WE89,WE00,CA00}. The effort spent to solve this (still open)
problem is justified since the geometry and the evolution of the Universe are
closely related to it~\cite{CS01,CA00,WI00}. 
Therefore, advances in this problem may lead us to a
better understanding of some of the crucial problems in both cosmology and
elementary particle physics. In the latter, this problem is related to another
not well understood  question: the mechanism of mass generation via the
spontaneous symmetry breaking (SSB), through scalar Higgs fields with a non-zero
vacuum expectation value (VEV). That mechanism also implies a new large
contribution to the cosmological term in the Einstein's equations~\cite{jdmv}. 
We named this the ``electroweak cosmological constant'' (ECC) in order to left
clear the difference with other quantum field contributions to the cosmological
constant as those mentioned before.

Independently of these theoretical issues, recent astronomical
observations~\cite{Pea99,FK98,Bea00,exp} strongly suggest a nonzero and positive
$\Lambda$. This result comes from observations of Type Ia
supernova~\cite{Pea99}, gravitational lensing frequencies for
quasars~\cite{FK98} and harmonics of cosmic background radiation (CBR)
anisotropies~\cite{Bea00}. These astronomical data gives $0.6 < \Omega_\Lambda <
0.8$, where $\Omega_\Lambda = \Lambda/3H^2_0$ is the $\Lambda$ density in units
of critical universe mass density, with $H_0 = 71$ km
s$^{-1}$Mpc$^{-1}$, (1 s = $1.52 \times 10^{24}$ GeV$^{-1}$ and 1
pc = $3.086 \times 10^{16}$ m~\cite{pdg}). Assuming $\Omega_\Lambda = 0.7$ we
have  
\begin{equation}
\Lambda_{\rm obs} =  4.8 \times 10^{-84}\; {\rm GeV}^2\quad  ({\rm or}\quad 
\Lambda_{\rm obs} G\approx10^{-122}),
\label{lambda}
\end{equation}
where we have used $G = 6.71 \times 10^{-39}$ GeV$^{-2}$ and $\hbar =  c = 1$. 

In this work we are interested only in the contributions of the scalar
Higgs sector to the vacuum energy density. Models with the scalar
fields, denoted collectively by $\Phi$, with the respective scalar potential
$V\left(\Phi\right)$, have a part in the energy momentum tensor given by  
\begin{equation} 
T_{\mu\nu} = \partial_\mu\Phi\partial_\nu\Phi -g_{\mu\nu}\left(\frac{1}{2} 
g^{\alpha\beta}\partial_\alpha\Phi\partial_\beta\Phi- V(\Phi)\right).  
\end{equation} 

The vacuum part in $T_{\mu\nu}$ is gotten when 
$\partial_\mu\Phi = 0$ so that, denoting the VEVs of the neutral component
of the Higgs fields collectively as $\Phi$ by $\langle\phi\rangle$, we have that
the vacuum energy momentum tensor is    
\begin{equation} 
T^{\rm vac}_{\mu\nu} = g_{\mu\nu}V\left(\langle\phi\rangle\right) =
g_{\mu\nu}\rho_{\Lambda_W} \equiv g_{\mu\nu}\frac{\Lambda_W}{8\pi G},  
\label{def} 
\end{equation}
where $\rho_{\Lambda_W}$ is the vacuum energy density. We can write the
purely electroweak contribution to the cosmological constant 
$\Lambda_W=8\pi GV(\langle\phi\rangle)$.

Hence, in addition to the intrinsic cosmological constant
in the Einstein equation, $\Lambda$ in Eq.~(\ref{eq}), there is a contribution
to the vacuum energy coming from SSB sector, so the observed
or effective cosmological constant is~\cite{indio}
\begin{equation}
\Lambda_{\rm eff}=\Lambda-8\pi G \vert V(\langle\phi\rangle)\vert.
\label{2cc}
\end{equation}

It is $\Lambda_{\rm eff}$ which should coincide with $\Lambda_{\rm obs}$ i.e.,
$\Lambda_{\rm eff}\approx \Lambda_{\rm obs}$, and this can be obtained if
\begin{equation}
\Lambda_{\rm eff}\approx \Lambda_{\rm obs}\quad(a);\quad
{\rm or}\quad \Lambda\approx \Lambda_{\rm obs},\quad 
8\pi G V(\langle\phi\rangle)=0 \quad (b). 
\label{finetuning}
\end{equation}

Eq.~(\ref{finetuning}a) implies that there exist a fine tuning between two
parameters which are different in the sense that
$\Lambda$ does not depend on the matter fields while $8\pi
GV(\langle\phi\rangle)$ does. We may wonder ourselves how this fine tuning is
possible between terms which so different physical origin. It will be more
natural that $\Lambda\approx\Lambda_{\rm  obs}$ while $8\pi G
V(\langle\phi\rangle)=0$. 
The first possibility, Eq.~(\ref{finetuning}a), can be implemented in any
electroweak model and for this reason it is not so much interesting as a
solution to the electroweak cosmological constant problem. On the other hand, it
is not clear at all in which kind of models (if any) the case in
Eq.~(\ref{finetuning}b) is realized. In other words, in general we will look for
an interval for the possible values for $\Lambda$ 
\begin{equation}
\Lambda_{\rm obs}+8\pi G\vert V(\langle\phi\rangle)_m\vert\leq \Lambda \leq 
\Lambda_{\rm obs}+8\pi G \vert V(\langle\phi\rangle)_M\vert
\label{verdadeira}
\end{equation}
in which $\vert V(\langle\phi\rangle)_M\vert$
($\vert V(\langle\phi\rangle)_m\vert$)
is the largest (smallest) value of $\vert V(\langle\phi\rangle)\vert$. 
It is well known, and we will review this in the next section, that in the
standard model $\vert V(\langle\phi\rangle)_M\vert= \vert 
V(\langle\phi\rangle)_m\vert \equiv
\vert V(\langle\phi\rangle)\vert$ and for this reason  only the fine
tuning~$\Lambda=\Lambda_{\rm obs}+8\pi G\vert V(\langle\phi\rangle)\vert$ is
allowed and $\Lambda_{\rm eff}\approx \Lambda_{\rm obs}$ as in
Eq.~(\ref{finetuning}a).    

Here a remark is in order. It is always possible to add an arbitrary constant
$V_0\equiv V(\Phi=0)$ to the scalar potential, in such a way that 
$V(\langle\phi\rangle)\to  V(\langle\phi\rangle)+V_0$ in Eq.~(\ref{2cc}). We
may consider $V_0$ as a bare cosmological constant in Einstein equation as in
Eq.~(\ref{eq}). This follows because if $L_{\rm matter}$ denotes the matter
Lagrangian, which includes the scalar potential $V(\Phi)$, we can also use
$L_{\rm matter}^\prime=L_{\rm matter}-V_0$ instead of just $L_{\rm matter}$.
Both Lagrangians give the same equation of motion for the 
matter, moreover, $V_0$ is a model independent parameter in the sense that it does
not depend on the fields of the model and it is the same for all electroweak
models. In fact, $V_0$ (or $\Lambda$) does not depend on the state of the
Universe. Hence, it is possible to identify $\Lambda$ with $8\pi G V_0$ i.e.,
$\Lambda \equiv 8\pi G V_0$ or, more properly that $\Lambda+8\pi G
V_0\to\Lambda^\prime$ and just omit the prime. 

We are also not assuming any form of exotic dark matter inducing a highly
negative pressure~\cite{lips}. Hence, we will neglect other
contributions to the vacuum energy density and consider only the contributions
of the Higgs scalars. It means that we will compare, under the same assumptions,
only the electroweak contributions to the vacuum energy density in several
electroweak models. 

The outline of this paper is the following. In Sec.~\ref{sec:sm} we 
consider the cosmological constant coming from SSB in the context of the
electroweak standard model. As it is known in that model a fine tuning,
only among the parameters of the scalar potential, is not allowed since it would
imply a rather light Higgs boson. Next, in Sec.~\ref{sec:multi}, we show that
the troubles survives in the multi-Higgs extensions of the standard model with
or without supersymmetry. In Sec.~\ref{sec:331} we analyze one of the so called
3-3-1 models, including 1-loop effective potential, in a version of the model
with only three Higgs triplets~\cite{PT93}. We also, briefly, comment the case
of the 3-3-1 model with a scalar sextet~\cite{331}. Our conclusions appear in
the last section.    

\section{The ECC in the standard model}
\label{sec:sm}

To clarify our goal in this work let us consider the ECC 
problem in the context of the standard electroweak model. There the
scalar potential is 
\begin{equation} 
V_{\rm SM} = \frac{1}{2}\left[\mu^2\Phi^\dagger\Phi +  \frac{\lambda}{2}
\left(\Phi^\dagger\Phi\right)^2\right],  
\label{pot}
\end{equation}
\noindent where $\mu^2 < 0$ and $\lambda > 0$ are parameters of the
potential and $\Phi = \left(\begin{array}{cc} \phi^+ & \phi^0
\end{array}\right)^T$ is the only Higgs scalar doublet present in the model. The
 neutral component $\phi^0$ gets a non-zero VEV $v_W\approx 246$ GeV. 
 When we take the potential at the minimum point we find   
\begin{equation} 
V_{\rm SM}(v_W) =  -\frac{\lambda}{4}\; v_W^4.  
\label{sm} 
\end{equation}  
implying that $\Lambda_W= 8\pi G V_{\rm SM}(v_W) \approx
-3.4\times 10^{-28}\;\lambda  \;{\rm GeV}^2$, 
which is, independently of the minus sign, several order of magnitude larger
than $\Lambda_{\rm obs}$ (see Eq.~(\ref{lambda})) unless we
impose that the constant of the scalar potential has an appropriate value:
$\lambda =1.4\times10^{-56}$. However, since the mass of the Higgs
scalar is given by $m_H = \sqrt{2\lambda}\,v_W$, 
that value for $\lambda$ implies a rather light scalar $m_H\sim
10^{-22}m_e$, where $m_e$ is the electron mass~\cite{jdmv}. A Higgs boson
with mass $\lesssim 5$ GeV has been ruled out by a variety of arguments derived
from rare decays, static properties like the magnetic moment of the muon, and
nuclear physics~\cite{GH90}. So, in the SM we cannot use a fine-tuning among its
parameters for obtaining a cosmological constant compatible with
the observed value, and at the same time to obtain a scalar Higgs field with
a mass of the order of 115.6 GeV~\cite{higgswg}. In other words, in the standard
model we have that $V_{\rm SM}(v_W)=-2\lambda\times10^9\,{\rm GeV}^4$.
Since $V_{\rm obs}=\Lambda_{\rm obs}/8\pi G=2.8\times10^{-47}\,{\rm GeV}^4$ it
means that $\vert V_{\rm SM}(v_W)/V_{\rm obs}\vert 
\approx\lambda\times10^{56}$ and since from the experimental data we
have that  $\lambda\stackrel{>}{\sim}  0.13$, we see that only the case in
Eq.~(\ref{finetuning}a) is possible in this model. 

From the phenomenological point of view this is not a 
fault of the (any) electroweak model. However, we may wonder ourselves 
if it is possible to built a model in which there is no a contribution to the
cosmological constant coming from the SSB sector and having, at the same time, a
realistic mass spectrum in the scalar sector. 

\section{Multi-Higgs models}
\label{sec:multi}  

In this section we will consider two popular extension of the standard model:
the two doublets, in Sec.~\ref{subsec:2h}, and the minimal supersymmetric
standard model (MSSM), in Sec.~\ref{subsec:mssm}.

\subsection{Two Higgs scalar non-supersymmetric model} 
\label{subsec:2h}  

In the two doublets non-supersymmetric extension of the standard model the Higgs
scalars are $\Phi_1 =\left(\begin{array}{cc} \phi_1^+ & \phi^0_1
\end{array}\right)^T$ and $\Phi_2 = 
\left(\begin{array}{cc} \phi_2^+ & \phi_2^0\end{array}\right)^T$  
with the VEVs $\langle\phi_1^0\rangle = v_1/\sqrt2$ and $\langle\phi_1^0\rangle = 
v_2/\sqrt2$ and $v^2_1+v^2_2=v^2_W$. Its $CP$-conserving scalar potential is given
by~\cite{GH90}   
\begin{eqnarray} 
V_{\rm TD} & = & \mu_1^2\,\Phi_1^\dagger\Phi_1 + \mu_2^2\,\Phi_2^\dagger\Phi_2
+ a_1\left(\Phi_1^\dagger\Phi_1\right)^2 + 
a_2\left(\Phi_2^\dagger\Phi_2\right)^2 + a_3\left(\Phi_1^\dagger
\Phi_1\right)\left(\Phi_2^\dagger\Phi_2\right)  \cr &+&
a_4 (\Phi_1^\dagger\Phi_2)(\Phi^\dagger_2\Phi_1)+ a_5\left[
\left(\Phi_1^\dagger\Phi_2\right)^2 + H.c.
\right],  
\label{pot2d}
\end{eqnarray} 
and we have assumed invariance under $\Phi_2\to-\Phi_2$~\cite{SH89}. 

Thus, the vanishing of the linear terms in the neutral fields gives the
constraints  
\begin{equation} 
-\mu_1^2 = a_1v_1^2 + \frac{1}{2}(a_3 + a_4+2a_5 )v_2^2, \qquad
-\mu_2^2  =  a_2v_2^2 +\frac{1}{2}(a_3 + a_4+2a_5)v_1^2,
\label{constrd}
\end{equation}
while the positivity of the
Higgs bosons mass matrices implies 
\begin{equation}
a_1,a_2>0,\;\;a_5<0,\;\; (a_3+a_4+2a_5)^2<4a_1a_2,\;\; a_4+2a_5<0.
\label{novos}
\end{equation}

Next, using Eqs.~(\ref{constrd}) the minimum of the potential is   
\begin{equation} 
V_{\rm TD}({\rm min})\equiv
V_{\rm TD}(v_1,v_2) = -\frac{1}{4}\,a_1v^4_1-\frac{1}{4}\,a_2v^4_2-
\frac{1}{4}\,(a_3+a_4+2a_5)v^2_1v^2_2.
\label{hm}
\end{equation}   
It is easy to verify that a solution to the equation
$V_{\rm TD}(v_1,v_2)=0$, that at the same time satisfies the constraints in
Eq.~(\ref{novos}), does not exist. Hence, in this model we have always a
negative contribution to the vacuum energy. In fact, we can write Eq.~(\ref{hm})
in terms of the masses of the neutral physical Higgs of the model $m_{H_{1,2}}$ 
\begin{equation}
V_{\rm TD}({\rm min})=-\frac{1}{16}\,
\frac{1}{a_1+a_2-(
a_3+a_4+2a_5)}\,\frac{m^2_{H_1}m^2_{H_2}}{\tilde{v}^2_1(v^2_W-\tilde{v}^2_1)}
\cdot v^4_W,
\label{twofinal}
\end{equation}
where $\tilde{v}_1$, is given by 
\begin{equation}
\tilde{v}^2_1=\frac{1}{2}\,\frac{2a_2-(a-3+a_4+2a_5)}{a_1+a_2-(a_3+a_4+2a_5)}
v^2_W,
\label{vtilde}
\end{equation}
has been obtained from the condition $\partial V_{\rm
TD}({\rm min})/\partial v_1=0$.

We see that in this two doublet
extension of the SM $V_{\rm TD}({\rm min})_M\approx V_{\rm TD}({\rm min})_m
\approx V_{\rm TD}({\rm min})$, and
the interval allowed for $\Lambda$ in Eq.~(\ref{verdadeira}) is a rather small
one i.e., in practice we have that like in the SM model 
$\Lambda\approx\Lambda_{\rm obs}+V_{\rm TD}({\rm min})$. Thus, as 
in the previous section the only fine tuning allowed is 
$\Lambda\stackrel{>}{\sim} +10^{56}\Lambda_{\rm obs}$. 
Notice also from Eq.~(\ref{twofinal}) that a vanishing contribution of the
scalar potential to the cosmological constant implies also a zero mass neutral
scalar. In this situation it is not worthing to calculate loop corrections to
the scalar potential.
 
\subsection{Minimal supersymmetric standard model} 
\label{subsec:mssm}  

The minimal supersymmetric model has also two scalar 
doublets, $H_1 = ( H^{0*}_1,\,  -H^-_1)^T$ 
and $H_2 = (H_2^+, \, H_2^0)^T$. The
associated scalar potential is in this case~\cite{MA97,haber2}
\begin{eqnarray}  
V_{\rm S} & = & m_1^2\vert H_1\vert^2 + m_2^2\vert H_2\vert^2 - m_3^2
\left(\epsilon_{ij}H^i_1H^j_2 + {\mbox{H. c.}}\right)  +
\frac{1}{8}\left[g^2\sum_i\vert H_1^\dagger\tau_i H_1 + H_2^\dagger \tau_i
H_2\vert^2 \right. \cr && \left. + g^{\prime2}\left(\vert H_1\vert^2 - \vert
H_2\vert^2\right)^2\right],  
\label{potsusy}
\end{eqnarray} 
where $m_1$, $m_2$ and $m_3$ are parameters with dimension of mass, $\tau_i$
$\left(i = 1, 2, 3 \right)$ are the Pauli matrices, $g^\prime =
g\tan{\theta_W}$ and $g = 2\sqrt{\sqrt{2}G_F}M_W$. As before we define the VEVs
as $\langle H^0_1\rangle = v_1/\sqrt2$, $\langle H^0_2\rangle = v_2/\sqrt2$,
with $v_1^2 + v_2^2 = v_W^2$, and the constraints equations are:
\begin{mathletters} 
\begin{eqnarray} 
-m_1^2 & = & \frac{1}{8v_1}\left[ v_1\left(g^2 + g^{\prime2}\right)\left(v_2^2 -
v_1^2\right)   + 4m_3^2v_2\right] , \\ -m_2^2 & = &
\frac{1}{8v_2}\left[v_2\left(g^2 + g^{\prime2}\right)\left(v_1^2 - v_2^2\right) 
 + 4m_3^2v_1\right].  
\end{eqnarray}
\end{mathletters} 

Therefore, taking the potential in the minimum we find 
\begin{eqnarray} 
V_{\rm S}({\rm min})\equiv
V_{\rm S}(v_1,v_2) &=& -\frac{g^2}{32\cos^2\theta_W}\left(v_1^2 -
v_2^2\right)^2,\nonumber \\ &=&
-\frac{v_1v_2}{8m^2_3}m^2_{H_1}m^2_{H_2}.  
\label{vs0} 
\end{eqnarray}  
In the second line $m^2_{H_{1,2}}$ denote, like in the previous model, 
the mass square of the two physical
neutral scalars of the model. The fine tuning $V_{\rm S}({\rm min})=0$ in
Eq.~(\ref{vs0}), occurring apparently with $v_1=v_2$, is not possible since it
implies that one of the neutral Higgs is massless, and we have a similar
situation to the two previously considered cases i.e., the minimum of
the potential is proportional to the mass(es) of the neutral Higgs scalar(s). On
the other hand, the 1-loop corrections vanishes because of the supersymmetry.
However, this occurs only in the early universe when $T>T_{\rm SUSY}$ and since
we are now at $T<T_{\rm SUSY}$ this is not a desirable
scenario. Anyway, the soft terms that have to be added to break the
supersymmetry will induce non-zero contributions to $\Lambda_W$, at both tree
and 1-loop level, which are also of the order of 
$\vert\Lambda_W\vert\stackrel{>}{\sim} 10^{56}\Lambda_{\rm obs}$, and this is
the only possibility.  
 
\section{The cosmological constant in a 3-3-1 model} 
\label{sec:331}  

Another model that we will consider here is the three scalar triplet version of 
the 3-3-1 model~\cite{PT93}. In this kind of model the gauge symmetry 
$SU(3)_C\otimes SU(2)_L\otimes U(1)_Y$ of the standard model  is
extended to $SU(3)_C\otimes SU(3)_L\otimes U(1)_N$. The  pattern
of symmetry breaking is $SU(3)_L\otimes U(1)_N$
$\stackrel{\langle\chi\rangle}{\longmapsto}SU(2)_L\otimes U(1)_Y$
$\stackrel{\langle\eta, \rho\rangle}{\longmapsto}U(1)_{\rm em}$ where  
\begin{equation}  
\eta =  \left(\begin{array}{c} \eta^0 \\  \eta_1^- \\  \eta_2^+
\end{array}\right), \quad  \rho =  \left(\begin{array}{c} \rho^+ \\  \rho^0 \\ 
\rho^{++} \end{array}\right), \quad  \chi = \left(\begin{array}{c} \chi^- \\ 
\chi^{--} \\ \chi^0 \end{array}\right),   
\label{trip}
\end{equation}\noindent  
transform under the $SU(3)_L\otimes U(1)_N$ group as $\left({\bf 3}, 0\right)$,
$\left({\bf 3}, 1\right)$ and $\left({\bf 3}, -1\right)$, respectively. The
neutral scalar fields develop the VEVs $\langle\eta^0\rangle = v_\eta/\sqrt2$, 
$\langle\rho^0\rangle =v_\rho/\sqrt2$ and $\langle\chi^0\rangle =
v_\chi/\sqrt2$, with $v_\eta^2 + v_\rho^2 = v^2_W$. According to the
symmetry breaking pattern we must have $v_\chi \gg  v_\eta, v_\rho$.  The
rest of the model representation content is given in Appendix~\ref{sec:a1}.    

\subsection{The tree level scalar potential} 
\label{subsec:331a}  

We take the more general scalar potential that is renormalizable and conserves
the total lepton number, i.e.,   
\begin{eqnarray} 
V_{\rm 331} & = & \mu_1^2\eta^\dagger\eta + \mu_2^2\rho^\dagger\rho + \mu_3^2
\chi^\dagger\chi + \lambda_1\left(\eta^\dagger\eta\right)^2 + 
\lambda_2\left(\rho^\dagger\rho\right)^2 + \lambda_3\left(\chi^\dagger\chi
\right)^2 + \left(\eta^\dagger\eta\right)\left[\lambda_4\left(\rho^\dagger
\rho\right) + \right. \cr  && \left.
\lambda_5\left(\chi^\dagger\chi\right)\right] + \lambda_6
\left(\rho^\dagger\rho\right)\left(\chi^\dagger\chi\right) +   
\lambda_7\left(\rho^\dagger\eta\right)\left(\eta^\dagger\rho\right) + 
\lambda_8\left(\chi^\dagger\eta\right)\left(\eta^\dagger\chi\right) +   
\lambda_9\left(\rho^\dagger\chi\right)\left(\chi^\dagger\rho\right) \cr &+& 
\frac{1}{2}\left(f\epsilon^{ijk}\eta_i\rho_j\chi_k + \mbox{H. c.}\right),  
\label{pot331} 
\end{eqnarray} 
where the $\lambda'$s are dimensionless constants and $\mu'$s and  $f$
have mass dimension~\cite{TO96}. As before, 
conditions for the extremum of $V_{331}$, besides the trivial solutions
$v_\eta=v_\rho=v_\chi=0$, imposes   
\begin{mathletters}
\begin{eqnarray} 
\mu^2_1+\lambda_1 v^2_\eta+\frac{\lambda_4}{2}v^2_\rho+
\frac{\lambda_5}{2}v^2_\chi+ \frac{A}{2\sqrt2  v^2_\eta} =0, \label{mus1}\\
\mu^2_2+\lambda_2 v^2_\rho+\frac{\lambda_4}{2} v^2_\eta
+\frac{\lambda_6}{2}v^2_\chi+ \frac{A}{2\sqrt2 v^2_\eta} = 0,  \label{mus2}\\  
\mu^2_3+\lambda_3v^2_\chi+\frac{\lambda_5}{2} v^2_\eta +\frac{\lambda_6}{2}
v^2_\rho+ \frac{A}{2\sqrt2 v^2_\chi}=0.
\label{mus3} 
\end{eqnarray} 
\label{mus} 
\end{mathletters} 
We have defined $A\equiv fv_\eta v_\rho v_\chi$ and we have assume all VEVs and
$f$ to be real. (Otherwise a physical phase remains in the model, which we can
choose as the phase of $v_\chi$, and we have CP violation~\cite{cp3}.) Unlike
the previous models the present one has two additional parameters with dimension
of mass: $v_\chi$ (which have to be larger than 246 GeV) and $f$, the coupling
constant in the trilinear term of the scalar potential, that in principle is an
arbitrary mass scale. 

The minimum of the full scalar potential is written as (up to the 1-loop order)
\begin{equation}
V_{331}({\rm min})\equiv V_{331}(v_\eta,v_\rho,v_\chi)=
V^{(0)}({\rm min})+ V^{(1)}({\rm min}),
\label{pottotal} 
\end{equation} 
where $V^{(0)}({\rm min})$ and $V^{(1)}({\rm min})$ denote the tree level and
the $1$-loop contributions ($\hbar=1$), respectively. We will consider the
possibility that the extra constraint $V_{331}({\rm min})=0$ depends only on the
parameters of the model. The tree level term is give by 
\begin{equation} 
V^{(0)}({\rm min})= -\frac{1}{4}\left(\lambda_1v_\eta^4 +\lambda_2v_\rho^4
\lambda_3v_\chi^4+ \lambda_4 v_\rho^2 v^2_\eta+
\lambda_5 v_\chi^2 v_\eta^2 +
\lambda_6 v_\chi^2 v_\rho^2+ \frac{f}{\sqrt2}\,
v_\eta v_\rho v_\chi\right),
\label{3310} 
\end{equation}
and we impose the extra constraint equation
\begin{equation} 
V^{(0)}({\rm min})=0, 
\label{treelevel} 
\end{equation} 
which implies that the scalar potential is flat at the tree level. 
A two dimensional projection of Eq.~(\ref{treelevel}) is shown in
Fig.~\ref{fig1}. 
In fact, Eq.~(\ref{treelevel}) defines a 3-sphere in the $V,v_\eta,v_\rho$
and $v_\chi$ space with $V=0$.
 
Before considering the 1-loop correction, we will show that, unlike the other
models considered above, in this 3-3-1 model there is a range of the 
parameters where the condition Eq.~(\ref{treelevel}) is
satisfied without spoiling the mass spectra of the model. In other words, after
the SSB the minimum of the scalar potential may be flat in all directions and
even in this case the model remains realistic. Of course, this is not a
prediction of the model since there are other ranges of the parameters where
this is not the case. 

From Eqs.~(\ref{3310}) and (\ref{treelevel}) we get
\begin{equation} 
f=-\frac{\sqrt2}{v_\eta v_\rho v_\chi}\left(\lambda_1v_\eta^4
+\lambda_2 v_\rho^4  +\lambda_3 v_\chi^4
+\lambda_4 v^2_\eta v_\rho^2+ \lambda_5 v^2_\eta v_\chi^2 
  + \lambda_6 v^2_\rho v_\chi^2 \right).  
\label{f} 
\end{equation}  
The condition $f<0$ is required by the positivity of the masses of the scalar
fields. 

Next, it is necessary to consider in detail the scalar mass
spectrum. In the scalar neutral sector we have the mass matrix  
\begin{equation} 
M^2_0=\left( \begin{array}{ccc} 2\lambda_1v^2_\eta-\frac{A}{2\sqrt2 v^2_\eta} &
\lambda_4v_\eta v_\rho+\frac{A}{2\sqrt2 v_\eta  v_\rho} & \lambda_5v_\eta
v_\chi+\frac{A}{2\sqrt2  v_\eta v_\chi}\\ & 2\lambda_2v^2_\rho-\frac{A}{2\sqrt2 
v^2_\rho} & \lambda_6v_\rho v_\chi+\frac{A}{2\sqrt2 v_\rho v_\chi}\\ & &
2\lambda_3 v^2_\chi-\frac{A}{2\sqrt2  v^2_\chi}  \end{array}\right).  
\label{m20}
\end{equation}

Besides the VEVs, there are six dimensionless free
parameters, $\lambda_{1,\cdots,6}$ in the mass matrix in Eq.~(\ref{m20}). 
We will use, just as an illustration, small
values for the $\lambda$'s in order to be sure that we are in a perturbative 
regime. Thus we use the following values for them:
\begin{equation}
\lambda_1 = 1.5,\;\lambda_2 =10^{-2} \;\;\lambda_3 = 10^{-5},\;
\lambda_4=10^{-2},\;
\lambda_5=10^{-3},\;\lambda_6=5\times10^{-2}, 
\label{para1}
\end{equation}
and $v_\eta=160$ GeV, $v_\chi=1.5$ TeV and $v^2_\rho=[(246)^2 -
v^2_\eta]\approx (186.86\,{\rm GeV})^2$. From Eq.~(\ref{f}) and (\ref{para1}) it
follows that  
\begin{equation}
f \approx-159\;{\rm  GeV}.
\label{fn}
\end{equation}
and the eigenvalues of the matrix in Eq.~(\ref{m20}) are (in GeV)
\begin{equation}
M_{H_1}\approx472,\; M_{H_2}\approx 160,\; M_{H_3}\approx 24.
\label{eigen}
\end{equation}

On the other hand, it is well known
that the final LEP's results indicate that the mass of the standard Higgs bosons
must be of about 115.6 GeV~\cite{lepfinal}. Although we know that
experimental bounds on the Higgs bosons mass are model dependent,
below we will show that in this sort of models there is not incompatibility with
the LEP data, since $\chi^0$ transforms as singlet under the $SU(2)_L\otimes
U(1)_Y$  symmetry, hence $H_3\approx \chi^0$ is almost singlet.  

First at all, we note that the symmetry eigenstates $\eta^0,\rho^0,\chi^0$
and the mass eigenstates $H_1,H_2$ and $H_3$ are related by an orthogonal 
matrix as follows
\begin{equation}
\left(\begin{array}{c}
\eta^0 \\ \rho^0 \\ \chi^0
\end{array}\right)\approx
\left(\begin{array}{ccc}
-0.8708 & 0.4882 & 0.0573\\
0.4889 & 0.8723& -0.0030\\
0.0507= & -0.0256 & 0.9983
\end{array}\right)
\left(\begin{array}{c}
H^0_1 \\ H^0_2 \\ H^0_3
\end{array}\right),
\label{mmatrix}
\end{equation}
and we recall that from Eq~.~(\ref{eigen}) $M_{H_1}>M_{H_2}>M_{H_3}$. This
``inverted'' mass spectrum is a consequence of imposing the extra constraint
equation in Eq.~(\ref{treelevel}), or Eq.~(\ref{f}). 

Notice that, in fact with the values of the parameters used above, 
$\chi^0$ is almost the lightest scalar: $\chi\approx.9983
H_3$. At LEP the Higgs boson of the standard model is supposed to be 
produced mainly via the Higgsstrahlung in $e^+e^-\to ZH$ in the $s$ channel
where the Higgs boson is radiated off an intermediated $Z$ boson. This process
depends on the trilinear interaction $ZZH$ which in the standard model (at the
tree level) is proportional to 
\begin{equation}
{\cal L}^{\rm SM}_{ZZH}=-\frac{g^{\prime\, 2}v_W}{\sin\theta_W}\,Z^\mu
Z_\mu H^0,
\label{zzh}
\end{equation}
where $v_W=246$ GeV and $\sin^2\theta_W(M_Z)=0.23113$~\cite{pdg}. On the
other hand, the scalar $\chi^0$ couples mainly with the exotic heavy fermion of
3-3-1 models. Thus, the signature of $e^+e^-\to
Z\chi$ decay is two $b$-jets, from the $Z$ decay, and large missing energy.
Besides, this process has a rather small cross section since, as we will show
below, the $\chi^0$ scalar is $Z$-phobic i.e., the coupling $ZZ\chi$ is
suppressed with respect to the $ZZH$ coupling in the standard model. Although we
have calculated the vertex $ZZ\chi$ exactly we will show below only the
expression in the approximation $v_\eta,v_\rho\ll v_\chi$. In the later case the
$ZZ\chi$-trilinear 
coupling is proportional to  
\begin{equation}
{\cal L}^{331}_{ZZ\chi}=-\frac{g^{\prime 2}\,v_W}{\sin\theta_W}
\,\left[\sin\theta_W\,\frac{v_\chi}{v_W}B^{\,2}\right]Z^\mu Z_\mu\chi^0,  
\label{zzc}
\end{equation}
where we have defined
\begin{equation}
B\approx\frac{\sqrt{1-4\sin^2\theta_W}}{\sin\theta_W\cos\theta_W}
\left[
\frac{v^2_\rho}{v^2_\chi}-\frac{1-4\sin^2\theta_W}{2\cos^2\theta_W}\,
\frac{v^2_\eta+v^2_\rho}{v^2_\chi}
\right].
\label{nova}
\end{equation}
With the value of $\sin\theta_W$ and the VEVs $v_\eta,v_\rho$ and $v_\chi$ used already
above we find  
\begin{equation}
\sin\theta_W\,\frac{v_\chi}{v_W}\,B^{\,2}\approx10^{-4}.
\label{oba}
\end{equation}
We see that the vertex $ZZ\chi$ is 0.01 per cent of the respective vertex of
the standard model given in Eq.~(\ref{zzh}) even if we assume that $\chi^0\equiv
H_3$. The $Z$-phobic character of $\chi^0$ is in fact
expected since this field transforms as a singlet under $SU(2)_L\otimes
U(1)_Y\subset SU(3)_L\otimes U(1)_N$ and it decouples from $Z^0$ when $v_\chi\to
\infty$. 

On the other hand, we see from Eq.~(\ref{mmatrix}) that the $\eta^0,\rho^0$ have
also a small components on the lightest scalar $H_3$, i.e.,
$\eta^0\sim\cdots0.0573H_3$ and $\rho^0\sim\cdots-0.0030H_3$, where the ellipse
denote the components in $H_{1,2}$. It means, respectively, a factor 
of 0.0103 and 0.008 in any cross section involving $\eta^0,\rho^0$.  
In this case, besides this suppression factor, the respective couplings with the
$Z^0$ have a factor $v_\eta/v_W\approx0.6504$ and $v_\rho/v_W\approx0.76$.
taking into account the mixing angles in Eq.(\ref{mmatrix}) we obtain that the
cross section $e^+e^-\to ZH_3$ through $\eta^0(\rho^0)$ the total suppression
factor $1.4\times10^{-3}$ and $5\times10^{-6}$. So, the LEP data applies only to
the heavy scalar $H_1$ and $H_2$. 

The massive pseudoscalar present in the model
has a mass 
\begin{equation} 
M^2_A=-\frac{A}{2\sqrt2}\left(\frac{1}{v^2_\eta}+
\frac{1}{v^2_\rho}+\frac{1}{v^2_\chi} \right), \label{ma1}  
\end{equation}
\label{mesca}
and, with the values of the parameters used above we have that
$M_A\approx414$ GeV. 

In the charged scalar sector we have the masses given by
\begin{equation} 
M^2_{++}=\frac{\lambda_9}{2}( v^2_\rho+v^2_\chi)-
\frac{A}{2\sqrt2}\left(\frac{1}{v^2_\rho}+ \frac{1}{v^2_\chi}\right), 
\label{m++}
\end{equation} 
for the doubly charged scalar, 
\begin{equation}
M^2_{+1}=\frac{\lambda_7}{2}(v^2_\eta+v^2_\rho)-
\frac{A}{2\sqrt2}\left(\frac{1}{v^2_\rho}+ \frac{1}{v^2_\chi}\right),\;
M^2_{+2}=\frac{\lambda_8}{2}(v^2_\eta+v^2_\chi)-
\frac{A}{2\sqrt2}\left(\frac{1}{v^2_\rho}+ \frac{1}{v^2_\chi}\right),\;  
\label{m+12} 
\end{equation} 
for the two singly charged scalars. Using $\lambda_7=\lambda_8=\lambda_9=0.1$
we obtain the numerical values $M_{++}\approx433
$ GeV, $M_{+1}\approx417$ GeV, and $M_{+2}\approx462$ GeV.

Of course, it is possible to obtain another set of values for the scalar Higgs 
masses by choosing another values for the dimensionless constants $\lambda$'s 
(recall also that $v_\chi \stackrel{<}{_{\sim}} 3.5$ TeV~\cite{JJ97}).  
A solution with all neutral scalar heavier than 114 GeV is
shown in the Appendix~\ref{sec:a3}).

We have shown that in this 3-3-1 model, at the tree level, it is possible to
have a realistic mass spectra while vanishing the contribution to the
cosmological constant. Next, we will show that this situation is stable when
radiative 1-loop corrections are included.

\subsection{Radiative corrections to the scalar potential} 
\label{subsec:331b}  

\begin{figure}[ht]
\begin{center}
\vglue -0.009cm
\mbox{\epsfig{file=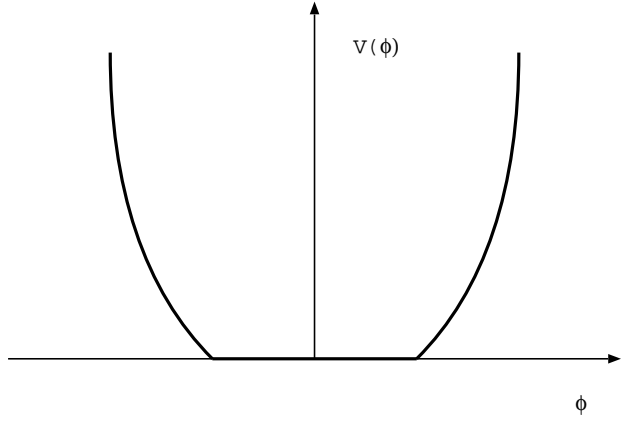,width=0.5\textwidth,angle=0}}       
\end{center}
\vglue 1cm
\caption{Two dimensional projection of the scalar potential in the 3-3-1 model. 
}
\label{fig1}
\end{figure}

Having showed that it is possible to get a flat potential at the tree level in a
3-3-1 model satisfying the condition in Eq.~(\ref{treelevel}), now we will be
concerned with the 1-loop radiative correction effects, and to see if at
this level the condition $V^{(1)}({\rm min})=0$ is possible. In other words,
even at the 1-loop level we have a potential as in Fig.~\ref{fig1}.

To calculate the effective potential at the 1-loop approximation we
appeal to one of the well known methods, see for instance
Refs.~\cite{colemanweinberg,jackiw}. In this approximation, all quantum
corrections can be extracted from  the quadratic part 
of the Lagrangian after shifting the neutral component of the scalars fields 
i.e., we shift the real part of these fields,
$\xi_\varphi\rightarrow\xi_\varphi+u_\varphi$, to calculate the first quantum
correction to the potential, i.e.,
$V=V^{(0)}+V^{(1)}$~\cite{jackiw}. According to 
this method, choosing a $R_\xi$ gauge with the Landau prescription~\cite{SH89},
a generic field $\varphi_\alpha$ give us the following 1 loop 
contributions to the scalar potential  
\begin{eqnarray} 
V_\alpha^{(1)}(u_{\varphi})&=&-\frac{i}{2}\int\frac{d^4k}{(2\pi)^4}\ln\hspace{0.1
cm}\det\left[i{{\cal{D}}^{-1}_{ab}(m_\alpha(u_\varphi),k)}
\right]\nonumber\\\nonumber\\ &=&-\frac{n_\alpha}{64\pi^2}\left[
\left(\frac{1}{\varepsilon}+\frac{3}{2}+\ln4\pi-
\gamma\right)m^4_\alpha(u_\varphi)-\frac{m^4_\alpha(u_\varphi)}{2}
\ln\frac{m^4_\alpha(u_\varphi)}{\sigma^4}\right],  
\label{Vi1} 
\end{eqnarray} 
where $n_\alpha$ stands for the respective degrees of freedom times +1 for
bosons and -1 for fermions,
$i{{\cal{D}}^{-1}_{ab}(m_\alpha(u_\varphi),k)}$  is 
the inverse of the propagator and $\sigma$ is an appropriate e\-ner\-gy scale.  The
second line in Eq.~(\ref{Vi1}) was evaluated using dimensional regularization
(see Appendix~\ref{a2}). To remove the infinities in Eq.~(\ref{Vi1}) we adopt
the minimal subtraction scheme $\overline{MS}$ in which all terms
proportional to $1/\varepsilon + 3/2 + \ln4\pi-\gamma$ are absorbed by
renormalization counter terms. It must be pointed out that the mass dimension
functions $m_\alpha(u_\varphi)$ above  are dependent on the $u_{\varphi}$, which
will be identified later with the real components of the neutral fields, and
only at the nontrivial minimum point it will assume the value of the physical
particle mass. Notice that there is no an infinite constant term, i.e., the
counter term to the cosmological constant, which we call $C_0$, which
determined by a renormalization condition, is finite  according to the
regularized integral above. This is a characteristic of this renormalization
scheme. Thus, the total contribution at the 1 loop order to the scalar
potential, for a model with a number $n$ of fields, is given by    
\begin{eqnarray} 
V^{(1)}=\frac{1}{64\pi^2}\sum_{\alpha=1}^n\frac{n_\alpha}{2}\,
m^4_\alpha(u_\varphi)
\hspace{0.05 cm}\ln\frac{m^4_\alpha(u_\varphi)}{\sigma^4}+C_0.  
\label{pot1} 
\end{eqnarray} 

This equation includes also a part due to the non-physical fields that give rise
to the Goldstone bosons. In fact, the functions $m^2_\alpha(u_\varphi)$ which
came  from the scalars fields, are the eigenvalues of the several mixing
matrices arising when the shift is realized in order to obtain loop
corrections. It is not difficult to see that when the constraints are imposed,
there is no contribution coming from these non-physical bosons for the effective
potential at the non trivial minimum. The reason for such a thing is that
Goldstone bosons are massless.        
The energy scale parameter $\sigma$ will be chosen at the scale
we discuss the new physics~\cite{bando}. Changing $\sigma$ does not affect
anything, since it is equivalent to a reparametrization of the coupling
constants. In our case, we will take $\sigma = 1.5$ TeV. This is due to the fact
that the main contribution to the effective scalar potential is given by the
heaviest particles present in the model and these  particles are expected to
have masses, in the 3-3-1 model, near that value. For the minimal
standard 
model, it is easy to see  that a natural choice is the vacuum expectation of the
Higgs field i.e., 246 GeV~\cite{SH89}. Here the situation is a little bit more
complicated since we have four physical neutral scalar fields. We have verified
that there is not a significative change in the final results for another
chooses of the energy scale, between 1 and 1.5 TeV, but maintaining the same
values for the coupling constants $\lambda_i$. 

Now that we have the effective potential up to 1 loop adding to the
tree level part given in Eq.~(\ref{3310}), we can fix the value of $C_0$ in
Eq.~(\ref{pot1}) by using the condition $V(0)=0$ and we have     
\begin{eqnarray} 
V_{331}({\rm min})= V^{(0)}(\xi_\varphi=u_\varphi)
+\frac{1}{64\pi^2}\sum_{\alpha=1}^nn_\alpha
\left[\frac{m^4_\alpha(u_\varphi)}{2}\hspace{0.05cm}
\ln\frac{m^4_\alpha(u_\varphi)}{\sigma^4}-\frac{m_\alpha^4(0)}{2}\hspace{0.05
cm} \ln\frac{m^4_\alpha(0)}{\sigma^4}\right],  
\label{V-1} 
\end{eqnarray} 
with $m^4_\alpha(0)$ meaning $m^4_\alpha$ computed at the trivial minimum  
i.e., at the origin. New constraints arise, and they differ from the previous
ones in Eq.~(\ref{mus}) only by corrections which came from
Eq.~(\ref{3310}), i e., we add functions $g_{v_\eta}$, $g_{v_\rho}$ and
$g_{v_\chi}$ to Eqs.~(\ref{mus1}), (\ref{mus2}) and (\ref{mus3}), respectively,
where $g_{v_\theta}$ stands for   
\begin{eqnarray} 
g_{v_\theta}=\left.\frac{1}{32\pi^2}\sum_{\alpha=1}^n n_\alpha m_\alpha^2(u_\varphi)
\left(\ln\frac{m^4_\alpha(u_\varphi)}{\sigma^4}+
1\right)\frac{dm^2_\alpha(u_\varphi)}{d u_\theta^2}
\right\vert_{u_\theta=v_\theta}.    
\end{eqnarray}  

\begin{figure}[ht]
\begin{center}
\vglue -0.009cm
\mbox{\epsfig{file=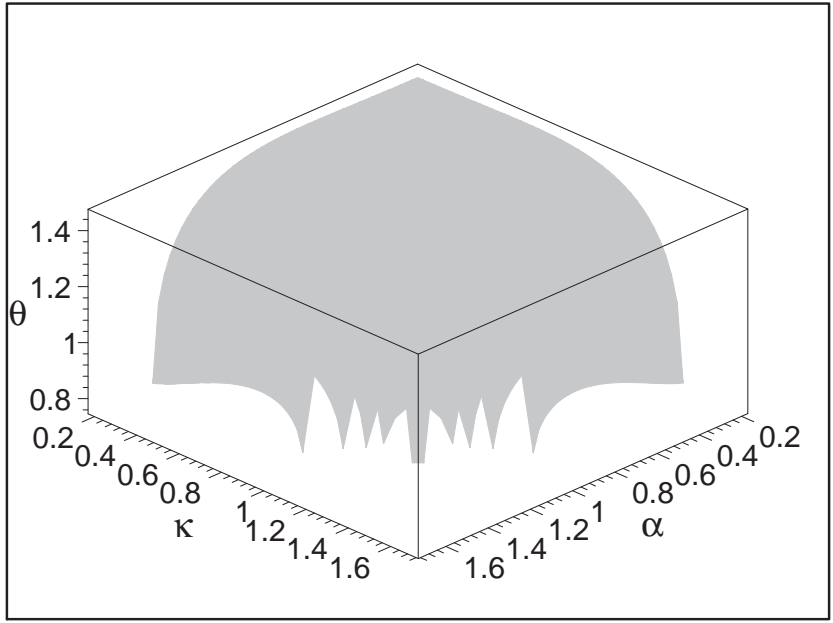,width=0.5\textwidth,angle=270}}       
\end{center}
\vglue 2cm
\caption{ Using Eq.~(\ref{V1TCOR}) we show the surface $V_{331}({\rm min})=0$
with as a function of the exotic quark masses. We have defined
$\alpha = m_{J}/v_\chi$, $\kappa = m_{j_1}/v_\chi$ and $\theta =
m_{j_2}/v_\chi$.}
\label{fig2}
\end{figure}

We will denote the potential $V$ at the minimum by $V_{331}{\rm (min)}$, and it is
given by    
\begin{eqnarray} 
V_{331}({\rm min})&=&-\frac{1}{64\pi^2}\left( \sum_{\varphi(phys)} n_\varphi
\left[(\ln\frac{m_\varphi^4}{\sigma^4}+1)m_\varphi^2\hspace{0.05
cm}v_\theta^2\frac{d\hspace{0.05 cm}m_\varphi^2}{d\hspace{0.05
cm}v_\theta^2}-\frac{m_\varphi^4}{2}\hspace{0.05 cm} 
\ln\frac{m_\varphi^4}{\sigma^4}\right]+3\sum_{i=1}^3{\mu_i^4}\hspace{0.05
cm}\ln\frac{\mu_i^4}{\sigma^4}\right),
\label{V1TCOR}  
\end{eqnarray} 
\noindent 
where $v_\theta^2(d/dv_\theta^2)=v_\eta^2(d/dv_\eta^2)+v_\rho^2(d/dv_\rho^2)
+v_\chi^2(d/dv_\chi^2)$, and $m^2_\varphi\equiv m^2_\varphi(v^2_\theta)$ denotes
the mass of the physical field $\varphi$. We  point out that this derivative
has to be done before the $\mu_{i}^2$ elimination.  The sum over $\varphi({\rm
phys})$ means that we are disregarding the non-physical fields in $V_{331}{\rm
(min)}$ according we have mentioned above. Hence, the mass dimension functions
$m_\varphi^2$ which appear in Eq.~(\ref{V1TCOR}), take in this point the value
of the mass of the respective particle $\varphi$. 
At this moment we have that the condition in Eq.~(\ref{finetuning}b) becomes
\begin{equation} 
V_{331}({\rm min})=0, 
\label{newc2} 
\end{equation} 
where $V_{331}({\rm mim})$ is given by Eq.~(\ref{V1TCOR}) and 
the main contribution comes from the heaviest
particles. They are the vector bosons $U^{++}$, $V^{+}$ and $Z^{\prime}$, three
exotic leptons $E_a$, and finally, the quarks $J$ and $j_i$. The contributions
of particles with masses below 500 GeV, as the top quark and the Higgs scalars
are negligible. We have also assumed that the heavy leptons are also in this
range. In Fig.~\ref{fig2} we show the surface satisfying the condition in
Eq.~(\ref{newc2}), using Eq.(\ref{V1TCOR}), as a function of the mass of the
exotic quarks and for $m_V = m_U = 1$ TeV. Notice that the upper limit of the
masses of the exotic quarks are near 1.44 TeV. 

Of course, there exist other values for the quark masses that satisfy the
condition of flat potential. This shows that in the 3-3-1 model context, unlike
the other electroweak models that we have considered above, the values of the
parameters stay in a reasonable range even when we impose a zero value
of the cosmological constant coming from the SSB of the model. 
Although we have considered the version of the model
with only three scalar triplets it is easy to convince yourselves that the same
will happen if we add a scalar sextet to the model as in Refs.~\cite{331}. 
In that kind of model there are at least two trilinear terms, and for this
reason it seems obvious that similar results should be obtained in this
extension of the model. Details will be given elsewhere.    

The radiative corrections will be inside of the validity domain of the
perturbation theory, only if $[a(\lambda)/ 64\pi^2]\;
\ln\,(m^2/\sigma^2)< 1$,
where $\lambda$ denotes the largest coupling constant of the electroweak
sector and $m$ is a generic mass parameter. It can be shown, using
the parameters given before, that this is the case for the 3-3-1 model. 

Unfortunately, in these models it is not
possible to obtain a tiny {\it and} positive $\Lambda_W$. At the tree level if
$V(\langle\phi\rangle)>0$, it means that the truly minimum of the potential is
at the origin $\langle\phi\rangle=0$ (see Fig.~\ref{fig1}). On the other hand,
at the 1-loop level if we impose that $V_{331}({\rm mim})>0$ using
Eq.~(\ref{V1TCOR}) or, that the fermion contributions dominate, the potential
becomes unbounded from below. Only negative or zero $\Lambda_W$ are possible to
be obtained. In fact, this argument is valid for any electroweak model. 

\section{Conclusions}
\label{sec:con}  

We conclude that, under reasonable hypotheses, in some gauge
models stringent bounds on symmetry breaking parameters come from the vacuum 
structure. In models with scalar potential as in Eq.~(\ref{pot}),
(\ref{pot2d}) or (\ref{potsusy}) 
the contribution of the SSB sector to the vacuum energy can be
made compatible with the observed value only after an extreme fine tuning
between two physically different parameters: a bare cosmological constant
$\Lambda$ and the purely SSB contribution $\Lambda_W=8\pi
GV(\langle\phi\rangle)$. This is the fine tuning indicated in
Eq.~(\ref{finetuning}a). In both models considered in Secs.~\ref{subsec:2h} and
\ref{subsec:mssm} we have $\vert V(\langle\phi\rangle)\vert_m\lesssim \vert
V(\langle\phi\rangle)\vert_M$ so, numerically there is no a large difference
with the situation of the standard model, thus from Eqs.~(\ref{verdadeira}),
(\ref{twofinal}) and (\ref{vs0}) we see that $\Lambda\approx \Lambda_{\rm
obs}+8\pi G \vert V(\langle\phi\rangle)\vert$, in those models.

However, in the 3-3-1 case, besides the fine tuning mentioned before, there
exist the possibility that the cosmological constant does not have an
electroweak contribution i.e., $\Lambda_W=0$, i.e., a fine tuning of the sort
of Eq.~(\ref{finetuning}b). In this model the lightest neutral scalar, $H_3$
($m_{H_3}\sim24 \,\mbox{GeV}\,\ll v_\chi\sim1.5$ TeV), 
may be almost singlet $H_3\approx \chi^0$.
This axion-like scalar may be a consequence of the Weinberg's no
go-theorem~\cite{WE89} valid in any adjustment mechanism approach to the
cosmological constant issue~\cite{sola}. 
As we mentioned before, and shown in Appendix~\ref{sec:a3}, it is possible for
another range of the parameters, to have an scalar $H_3$ with a mass of the
order of 156 GeV or even a little bit larger ($\approx200$ GeV). However, even 
in this case we have that $\chi^0\approx H_3$ and $m_{H_3}\ll v_\chi$. 

The Yukawa interactions in this 3-3-1 model are
given by (see the notation in Appendix~\ref{sec:a1}) 
\begin{equation}
-{\cal L}_\chi=
G^j_{ik}\overline{Q}_{iL}j_{kR}\chi ^{*}+
G^J\;\overline{Q}_{3L}J_{R}\chi+
G^E_{ab}\;\overline{(\Psi) }_{aL}E_{bR}\chi +H.c.,
\label{yu}
\end{equation}
where all fields in Eq.~(\ref{yu}) are symmetry eigenstates. We see that
$\chi^0$ couples only exotic heavy quarks and leptons with the known 
particles, so the decays $\chi^0\to q\bar{J},q\bar{j},E^+l^-$ are kinematically
forbidden; $\chi^0\to ZZ$ and $\chi^0\to Z^\prime Z^\prime$ are also 
energetically forbidden since always $M_{Z^\prime}>m_{H_i}$, and if 
$m_{H_3}<2M_Z$. Hence, for a range of the parameters, $H_3$ is a stable heavy
neutral scalar and a good candidate for cold dark matter. 

Due to the suppression factor shown in
Eq.~(\ref{oba}) even if $m_{H_3}\approx 200$ GeV, as in Appendix~\ref{sec:a3},
the decay $H_3\to ZZ$, which is now energetically allowed, has a width
$\Gamma(H_3\to ZZ)\approx3\times10^{-8}$ MeV which, although is small enough to
have consequences in accelerator physics, it is larger than the
width of the universe ($\sim10^{-39}$ MeV) i.e., the mean life $\tau_{H_3}$
is $\sim10^{-31}$ times the age of the universe.

The effective potential must be flat order by order and,
since the loop corrections depend on the masses of the other particles (exotic
fermions and vector bosons), any fine tune implies constraints on the masses of these
particles too. 
In particular, we show in Fig.~\ref{fig2} that if the masses of the
vector bosons are of the order of 1 TeV, the flatness condition of the effective
potential at the 1 loop level implies constraints on the masses of the exotic
quarks of the model which are compatible with what is expected on
phenomenological grounds~\cite{das}. With solutions to eigenvalues of the mass
matrix in Eq.~(\ref{m20}) in which 
the scalar $H_3$ has a mass of the order of 150 GeV our analysis
summarized up in Fig.~2 has to be redone, allowing in this case exotic quarks
with masses larger than the values appearing in the figure. 

From the phenomenological point of view, the fact that
the contribution to the vacuum energy is large is not an argument against
any electroweak model, however it is certainly a virtue of 3-3-1 models, and
probably other models with trilinear scalar interactions, that for a domain of
the parameters, the scalar potential is flat in all
directions and, at the same time, but it has still a realistic Higgs scalar mass
spectra. We recall that in 3-3-1 models trilinear terms are necessary 
since only with them the gauge symmetry of the scalar potential is just
$SU(3)_L$ and not, say, $[SU(3)_L]^3$ i..e, these interactions are necessary for
having the correct number of Goldstone bosons. Incidentally, we would like to
point out that no one of the proposed solutions to the observed cosmological
constant using scalar fields, have introduced trilinear
interactions~\cite{vilenkin}. Moreover, these ideas may be implemented also in
inflation scenarios~\cite{ag}.    

Summarizing, in the case of 3-3-1 models the fine tuning is as in 
Eq.~(\ref{finetuning}b) or, since we have $V(\langle\phi\rangle)_m=0$
in Eq.~(\ref{verdadeira}),
\begin{equation}
\Lambda_{\rm obs}\leq \Lambda\leq \Lambda_{\rm obs}+8\pi G\vert 
V(\langle\phi\rangle)_M\vert.
\label{mesmo}
\end{equation}
It is the possibility that $\Lambda=\Lambda_{\rm obs}$ that is interesting. In
this case $\Lambda$ can be considered a genuine gravitational parameter like the
Newton constant, $G$.
Finally, we would like to say that since $\langle\phi\rangle$, and hence
$V(\langle\phi\rangle)$, varies during the evolution of the universe, the value
of $\Lambda_W$ also varies with $T$. At early epoch of the Universe the effects
of temperature will change the value of the minimum of the potential by several
order of magnitude, and $\rho_\Lambda(T)=8\pi G V_{331}({\rm min})(T)$
will also change with the temperature. It would be interesting if in the 3-3-1
models  there exist a domain of the parameters in which the broken symmetry is
not restored at high temperature~\cite{senjanovic}.

\acknowledgements   

This work was supported by Funda\c{c}\~ao de Amparo \`a Pesquisa do Estado de
S\~ao Paulo (FAPESP), Conselho Nacional de  Ci\^encia e Tecnologia (CNPq) and by
Programa de Apoio a N\'ucleos de Excel\^encia (PRONEX). 
One of us (M. D. T.) would like to thank the Instituto de F\'\i sica Te\'orica
of the UNESP for the use of its facilities. The authors would like to thanks G.
E. A.  Matsas for useful discussions.      

\newpage 

\begin{appendix}  

\section{The 3-3-1 Model.} 
\label{sec:a1}  

The fermions of the 3-3-1 model that we have considered in this work are the
following~\cite{PT93}
\begin{mathletters} 
\label{fermions}
\begin{eqnarray}  
\Psi_{aL} & = & \left(\begin{array}{c}\nu_a \\ l^-_a \\ E^+_a
\end{array}\right)_L \sim \left({\bf 3}, 0\right),\;a=e,\mu,\tau
 \label{lep} \\  Q_{iL} & =
& \left(\begin{array}{c} d_i \\ u_i \\ j_i \end{array}\right)_L
\sim \left({\bf 3}^*, -\frac{1}{3}\right),\;1=1,2;  \label{quark1} \qquad Q_{3 L} =
\left(\begin{array}{c} u_3\\ d_3\\J
\end{array}\right)_L \sim \left({\bf 3}, \frac{2}{3}\right).
\label{quark2}   
\end{eqnarray} 

\noindent The left-handed
fermion fields have their right-handed counterparts transforming as singlets of
$SU(3)_L$ group, {\it i. e.},   
\begin{eqnarray}  
\lefteqn{l^-_{aR} \sim \left({\bf 1}, -1\right), \quad E^+_{aR} 
\sim \left({\bf 1}, +1\right),}  \\ && 
U_R \sim \left({\bf 1}, 2/3\right), \quad D_R \sim \left({\bf 1},
-1/3\right), \quad   J \sim \left({\bf 1}, 5/3\right), \quad j_{1,2R} \sim
\left({\bf 1}, - 4/3\right).
\label{quarkr}
\end{eqnarray}
\end{mathletters}\noindent 
In Eqs. (\ref{fermions}) the numbers 0, 2/3, $-$1/3, 5/3, $-$4/3 and $\pm$1 are
the $U(1)_N$ charges. Here we are defining $U_R = u_R, c_R,t_R$ and $D_R = d_R,
s_R, b_R$. In order to avoid anomalies one of the quark families must transform
in a different way with respect to others. 

In the gauge boson sector the single charged
$\left(V^\pm\right)$ and double charged $\left(U^{\pm\pm}\right)$ vector
bileptons, together with a new neutral gauge boson $Z^{0\prime}$, complete the
particle spectrum with the charged $W^\pm$ and the neutral $Z^0$  gauge bosons
from the SM. The content of the scalar sector is the three triplets of the Eqs.
(\ref{trip}). 

As we said before, in a version of the model where there is no heavy leptons,
one additional sextet have to be added~\cite{331}. However, since the sextet is
introduced to gives mass to the known charged leptons the respective VEVs are
smaller than the VEVs of the other Higgs scalars. Thus, its contributions to the
vacuum energy may be negligible.  

\section{Heavy scalars solutions}
\label{sec:a3} 

It is possible to obtain another set of values for the scalar masses
by choosing another values for the dimensionless constants.  

\begin{equation}
\lambda_1 = 1.5,\;\lambda_2 = 0.1,\;\;\lambda_3 = 10^{-3},\;\lambda_4=1.5,\;
\lambda_5=0.5,\;\lambda_6=1.5, 
\label{a31}
\end{equation}
and with the VEVs with same values used in Sec.~\ref{subsec:331a}, from
Eq.~(\ref{f}) and (\ref{para1}) it follows that  
\begin{equation}
f \approx-4861\;{\rm  GeV}.
\label{a32}
\end{equation}
and the eigenvalues (in GeV) of the matrix in Eq.~(\ref{m20}) are in this case
\begin{equation}
M_{H_1}\approx2.3\times10^3,\; M_{H_2}\approx 293,\; M_{H_3}\approx 156.
\label{a33}
\end{equation}

The symmetry eigenstates $\eta^0,\rho^0,\chi^0$
and the mass eigenstates $H_1,H_2$ and $H_3$ are now related by the orthogonal 
matrix 
\begin{equation}
\left(\begin{array}{c}
\eta^0 \\ \rho^0 \\ \chi^0
\end{array}\right)\approx
\left(\begin{array}{ccc}
-0.7640 & -0.6045 & -0.2253\\
0.6434 & -0.7397 & -0.1969\\
0.0476 & 0.2954 & -0.9542
\end{array}\right)
\left(\begin{array}{c}
H^0_1 \\ H^0_2 \\ H^0_3
\end{array}\right),
\label{a34}
\end{equation}
and we see that also in this situation the $\chi^0$ is almost the lightest
scalar $H_3$. We have also found solutions with $m_{H_3}\approx200$ GeV,
however, the respective $\lambda$'s in Eq.~(\ref{a31}), are dangerously near
at the perturbative limit of the theory. For latter reason we expect that
$m_{H_3}<2M_z$ solutions are favored for a range of the dimensionless coupling
constant as in Eqs.~(\ref{para1}) and (\ref{a31}).

As we said in Sec.~\ref{sec:con}, $\chi^0$ 
is mainly a stable neutral scalar $H_3$ and, if $m_{H_3}<2M_Z$ it may
may be a good candidate for cold dark matter.

\section{One loop correction to the potential}
\label{a2} 

In the 1-loop approximation, the
whole $\hbar$ contribution for the effective potential can be extracted from the
quadratic part of the Lagrangian \cite{jackiw}. It is, for computational
convenience, chosen a gauge fixing of $R_\xi$ type, see Ref.~\cite{SH89}. The
basic integral that must be computed, using dimensional regularization, is given
by~\cite{miran}    
\begin{eqnarray} 
V^{(1)}&=&-\frac{i\hbar}{2}\int \frac{d^4k}{(2\pi)^4} \ln \hspace{0.1 cm}\det
\hspace{0.1 cm} D_{\alpha\beta} (k,m)\nonumber\\ \nonumber\\ 
&=&-\frac{\hbar\sigma^{2\epsilon}}{2} \int \frac{d^Dk}{(2\pi)^D}
\ln(k^2-m^2)\nonumber\\\nonumber\\ 
&=&\frac{-i\hbar\sigma^{2\epsilon}}{2(4\pi)^{\frac{D}{2}}}\Gamma(-D/2)\hspace{0.1
cm}m^D\nonumber\\\nonumber\\  &=&-\frac{\hbar}{32\pi^2}m^4
\left[\frac{4\pi\sigma^2}{m^2}
\right]^\epsilon\Gamma(-2+\epsilon)\nonumber\\\nonumber\\ 
&=&-\frac{1\hbar}{64\pi^2}m^4\left
[\frac{1}{\epsilon}+\frac{3}{2}-\gamma+\ln4\pi-\ln\frac{m^2}{\sigma^2}\right],  
\end{eqnarray}   
where $\sigma$ is a  mass scale parameter introduced to keep the correct
dimension for $V^{(1)}$ when the integral is extended to others dimensions. And
$\epsilon=(4-D)/2$. It was also done the usual expansion for the $\Gamma$
function and taken the limit $D\rightarrow 4$, where it was possible. The
formula above must be multiplied by the number of the degrees of freedom of the
respective field, including a minus one factor for fermionic fields, which gives
a nonzero contribution to the effective potential. These numbers are, 6 for a
charged vectorial boson, 3 for a neutral vectorial boson, 2 for a charged
scalar, 1 for a neutral scalar and -4 for a lepton and -12 for a quark. 
The mass parameter, $m$, in $V^{(1)}$ are functions of the shifts in the real
components of the neutral scalars fields which get a vacuum expectation value
\cite{colemanweinberg,jackiw}. We call these shifts by
$u_\varphi\equiv(u_\eta,u_\rho,u_\chi)$, and are to be identified at the end of
the computation with their respective fields. Renormalization guarantees that we
can absorb the infinities through  redefinition of the arbitrary parameters of
the model. We use a renormalization prescription based in a modified minimal
subtraction of the kind $\overline{MS}$~\cite{bardeen} in the sense that we will
keep only the logarithmic term as loop corrections. Thus the whole one loop
corrections is summarized in    
\begin{eqnarray} 
V^{(1)}=\frac{\hbar}{64\pi^2}\sum^nn_\alpha m^4_\alpha(u_\varphi) 
\ln\frac{m^2_\alpha(u_\varphi)}{\sigma^2},   
\end{eqnarray} 
where $n_\alpha$ stands for the number related to the degrees of freedom, as we
said before, and the sum is over the $n$  physical and non-physical fields of
the theory.  
\end{appendix}

\vglue 0.01cm

\vglue 0.01cm

\end{document}